\documentstyle [epsfig] {l-aa}

\newif\iffigure

\begin{document}

\thesaurus{03(11.09.1; 11.09.4; 11.11.1; 11.19.3)}
 
\title{Evidences for an expanding shell in the Blue Compact Dwarf Galaxy
 Haro~2}

\author{F. Legrand \inst{1}, D. Kunth \inst{1}, J.M. Mas-Hesse \inst{2}
\and J. Lequeux \inst{3}}
\offprints{F. Legrand}
\institute{
  Institut d'Astrophysique de Paris, CNRS, 98bis boulevard Arago, F-75014 
  Paris, France.
  \and
  LAEFF--INTA, Apdo 50727, E-28080 Madrid,
  Spain.
  \and
  Observatoire de Paris, DEMIRM , 77, Av. Denfert Rochereau, F-75014 Paris,
  France. 
}

\date{received 08.01.1997 ; accepted 28.04.97 }

\maketitle
\markboth{ F. Legrand et al.: Evidences of an expanding shell in the
Blue Compact Dwarf Galaxy Haro~2 }{  }
\begin{abstract}

Long-slit observations of the blue compact galaxy Haro~2 have been
performed around H$\alpha$ and H$\beta$. The main aim of these observations
was to detect the H$\alpha$ emission originating in the partially ionized
wind outflowing at 200 km/s, that had been previously detected with the
Hubble Space Telescope (HST).   A shallow broadening of the H$\alpha$ line
wings has been observed, consistent with the existence of an expanding
shell. The rotation curve shows two dips at the same systemic velocity as
the nucleus. We interpret this feature as an evidence that the expanding
shell is decoupled from the disk rotation. At the positions of the dips the
H$\alpha$ line is clearly broadened with respect to the central core. This
broadening is produced by the outer layers of the expanding shell.  From
the position of these dips we estimate the size of the shell to be around
20'' in diameter, with a corresponding kinematical age between 5 and 6
Myr. This shell has most certainly been powered by the massive star
formation process that takes place in the central region of this galaxy. A
comparison of the H$\alpha$ and Ly$\alpha$ profiles shows that Ly$\alpha$
is significantly broader than H$\alpha$, with an additional emission in the
red wing.  We interpret this redshifted source of Ly$\alpha$ emission as
line photons backscattered by the receding part of the expanding shell.
These observations outline the extremely high sensitivity of the Ly$\alpha$
line to the structure and kinematics of the interstellar medium (ISM). 
Finally the analysis of 
stellar Balmer lines in the H$\beta$ region indicates that stars less 
massive than 10 M$\odot$ have probably been formed.
 
  \keywords{Galaxies: ISM --
             Galaxies: starburst -- 
             Galaxies: kinematics and dynamics --
             Galaxies: individual: Haro 2}
\end{abstract}
 

\section{ Introduction }

Theoretical studies of primeval galaxies (Partridge \& Peebles 1967, Meier
1976) have suggested that they could initially experience a strong massive
star formation episode, and should therefore exhibit strong Ly$\alpha$
emission. If so this line would be an ideal tracer of primeval galaxies at
high redshifts.  Since Blue Compact Dwarf Galaxies (BCD) experience also
strong star formation events, they could be regarded as local counterparts
of the primeval galaxies, hence constituting an excellent laboratory to
study the relation between star formation and Ly$\alpha$ emission (Meier
\& Terlevich 1981, Hartmann et al. 1988).  The first observational results
in studying Ly$\alpha$ emission in HII galaxies were nevertheless
surprising, since the Ly$\alpha$ emission came out much fainter than
expected in some starburst galaxies, while no emission at all could be
detected in others.

The problem of Ly$\alpha$ emission in galaxies experiencing starburst
episodes has been studied by numerous authors throught the last years (Neufeld
1991, Charlot \& Fall 1991, Terlevich et al. 1993, Calzetti et al. 1994,
Kunth et al. 1996, Giavalisco et al. 1996) and its interpretation has
evolved with time. The first IUE (International Ultraviolet Explorer)
studies suggested a correlation between the presence of Ly$\alpha$ emission
and metallicity, hence dust alone was first pointed out as the principal
cause for the reduction of Ly$\alpha$ emission.  New results from HST
(Kunth et al. 1994) have shown that galaxies with very low metallicity, as
IZw18, can have no Ly$\alpha$ emission while some
galaxies with more metals, like Haro~2 (Lequeux et al. 1995), exhibit
Ly$\alpha$ emission. On the other hand the spectral resolving power of the
Goddard High Resolution Spectrograph onboard HST made possible to detect
P~Cygni profiles in the Ly$\alpha$ line of several starburst galaxies
(Kunth et al. 1996).  This suggest that the detectability of the line is
strongly affected by the geometrical and kinematical properties of the
interstellar medium.  In BCD's, massive stars are numerous and power strong
stellar winds that have a significant impact into the surrounding
interstellar medium.  Recent observations (Marlowe et al. 1995, Martin
1996, Izotov et al. 1996, Yang et al. 1996) have revealed the presence of
winds, shells and bubbles expanding at high velocities in a number of
starburst galaxies. These objects are therefore well suited to investigate
the link between the Ly$\alpha$ emission and the properties of the ISM.

In this context, HST observations of a sample of emission-line galaxies
(Kunth et al. 1996) around Ly$\alpha$ and OI 1302~\AA\ have been
performed. Among them Haro~2 appears to be a relatively metal-rich blue
compact galaxy with a metallicity around $\frac{1}{3}$ solar (at least 10
times more than IZw~18).  Its recession velocity is 1465 km/s (Lequeux et al.
1995) and its absolute magnitude, assuming H=70 km/s/Mpc is $M_{B}=-18.7 $
(Kunth \& Joubert 1985). The apparent extinction in this galaxy varies from
E(B-V)=0.12 from the far-UV (Mas-Hesse \& Kunth 1997, in prep.) to E(B-V)=0.7 
from the Balmer decrement (Davidge 1989) and E(B-V)=1.4 using the H$\beta$ 
over Brackett $\gamma$ ratio (Davidge \& Maillard 1990).  This discrepancy
between the extinctions measured at different wavelengths is a frequent
property of extragalactic HII regions which is attributed to
inhomogeneities in the foreground absorbing screen (Lequeux et al.
1981). At short wavelengths, radiation emerges preferentially through
places with low extinction; consequently the extinction measured at these
wavelengths appears smaller than the average extinction which is better
measured at longer wavelengths.  Lequeux et al. (1995) observations have
shown that the Ly$\alpha$ emission is clearly asymmetric, with a clear
P~Cyg profile revealing the existence of a partially ionized gas, outflowing
at 200 km/s from the main central HII region.  The interpretation suggested
 by Lequeux et al. (1995) is that a neutral expanding medium is
responsible for the Ly$\alpha$ absorption.  Since absorption features due
to OI 1302~\AA\ and SiII 1304~\AA\ seem to be centered at the velocity of
the outflowing partially ionized gas (200 km/s), the neutral expanding medium
should be at this velocity, pushed by the ionized expanding medium and
absorbing the blue part of the Ly$\alpha$ emission line. Moreover, the red
wing of the observed Ly$\alpha$ emission is broad suggesting the presence
of an additional source of photons which could be due to the emission of
the receding ionized medium or to backscattering of Ly$\alpha$ photons on
the receding neutral gas.

The objective of this work is the analysis of the Haro~2 H$\alpha$ emission
line profile
in order to detect the broad component potentially emitted by this ionized
expanding medium, and to compare this profile with that of Ly$\alpha$.
Since these two lines are produced by the same physical process, hence
essentially in the same regions, the differences in their profiles reflect
the process which affects only the Ly$\alpha$ line, mostly the resonant
diffusion in the neutral gas. This will allow us to pin down to what
extent the structure of the ISM is influencing the profile of the
Ly$\alpha$ line.

We present the optical observations in Sect.~2. In Sect.~3 we discuss the
shape of the H$\alpha$ and Ly$\alpha$ profiles. The stellar population and
its evolutionary state is analyzed in Sect.~4. Finally, we discuss the
results in Sect.~5.


\section{ Observations and data reduction }

Observations were made using the ISIS double arm spectrometer at the 4.2~m
William Herschel Telescope in La Palma on February 27, 1995.  A long slit,
1.5 arcsec wide, was positioned over the central part of Haro~2 (RA=10~h
29~m 22~s, DEC=54$\degr$ 39$\arcmin$ 29$\arcsec$) with a position angle of
45$\degr$. The spatial resolution was 0.35 arcsec/pix. Exposures of 1800~s
were performed in the spectral ranges 3630-5150 \AA \ in the blue, and
6410-6810 \AA \ in the red, with respective spectral dispersions of 1.54
\AA/pix and 0.41 \AA/pix. The reduction of the images were made using the
standard procedures of IRAF. After wavelength calibration we have further
corrected for a slight systematic offset (1 pixel) in the direction of
dispersion between the line positions of the red and the blue spectra.
This could be due to a small mechanical flexure of the arms of ISIS during
the observations.

The seeing was unfortunately very poor during these observations, giving a
FWHM for stellar images around 5$\farcs$5. We have compared this data with
some other observations of Haro~2 obtained in March 1995 with the HEXAFLEX
multifiber device in La Palma (in collaboration with S. Arribas and
E. Mediavilla), in good seeing conditions (around 1$\arcsec$) to estimate to
which extent the bad seeing had degraded the spatial resolution of our
spectroscopic observations. We  conclude that the effect of the seeing
is mostly a ``smoothing'' of the spatial variations. We have analyzed
H$\alpha$ at every pixel along the slit (steps of 0$\farcs$35) using bins
of 1$\farcs$75 in order to study the general trend in the spatial behaviour
of its FWHM and to construct the rotation curve for Haro~2.


\section{Lines profiles}

We have first measured the positions and widths of all the emission lines
for 3 slit positions, respectively at the center (CE), defined as the
position of maximum brightness of the line emission, at 7'' from the center
in the north-east direction (NE) and at 7'' from the center in the
south-west direction (SW).  The results are shown in Table~\ref{tab:mesures}.

    \begin{table}
    \small
    \caption [] {Haro~2 line measurement} \label{tab:mesures}
    \begin{flushleft}
    \begin{tabular}{ccccccc}
    \hline
Line & Position & $\lambda_{obs}$ & $\lambda_{lab}$ & $\sigma_{obs}$ & $\sigma_{dec}$  & $V_{rec}$\\
    \hline
    (1) & (2) & (3) & (4) & (5) & (6) & (7)\\
    \hline
NII$\lambda$6548	&	NE	&	6579.6	&	6583.4	&	66.3	&	63.41	&	1450	\\
NII	&	CE	&	6579.4	&	6583.4	&	51.4	&	47.56	&	1433	\\
NII	&	SW	&	6578.9	&	6583.4	&	63.7	&	60.7	&	1413	\\
H$\alpha$&	NE	&	6594.5	&	6562.9	&	58.5	&	55.2	&	1444	\\
H$\alpha$&	CE	&	6594.1	&	6562.9	&	54.4	&	50.8	&	1428	\\
H$\alpha$&	SW	&	6593.9	&	6562.9	&	59.2	&	55.9	&	1418	\\
NII$\lambda$6584	&	NE	&	6615.3	&	6583.4	&	55.8	&	52.4	&	1454	\\
NII	&	CE	&	6614.9	&	6583.4	&	53.3	&	49.7	&	1436	\\
NII	&	SW	&	6614.6	&	6583.4	&	58.7	&	55.4	&	1425	\\
SII$\lambda$6717	&	NE	&	6749.1	&	6716.4	&	56.7	&	53.5	&	1461	\\
SII	&	CE	&	6748.7	&	6716.4	&	55.4	&	52.1	&	1442	\\
SII	&	SW	&	6748.1	&	6716.4	&	55.5	&	52.1	&	1416	\\
SII$\lambda$6731	&	NE	&	6763.6	&	6730.8	&	55.4	&	52.1	&	1464	\\
SII	&	CE	&	6763.1	&	6730.8	&	54.9	&	51.5	&	1443	\\
SII	&	SW	&	6762.7	&	6730.8	&	55.3	&	51.9	&	1421	\\
    \hline
    \end{tabular}
    \end{flushleft}
{\em Columns: }

(1) : Line identification. \\
(2) : SW, CE and NE respectively for the SW, central and  NE regions (integrated over 7'').\\
(3) : Observed wavelength in \AA.\\
(4) : Rest wavelength in \AA.\\
(5) : Measured $\sigma$ parameter in km/s. \\
(6) : The $\sigma$ parameter deconvolved from the instrumental profile (1 \AA \ resolution), in km/s. \\
(7) : Recession velocity in km/s.\\
    \end{table}

Since the FWHM of the calibration lines is 1 \AA \ in the red (3 times less
than the H$\alpha$ FWHM), the H$\alpha$ profile is well resolved. We have
computed a deconvolved $\sigma = \frac{FWHM}{2\sqrt{2 ln(2)}}$ parameter
for H$\alpha$, which is respectively 55, 51, 55 km/s (+/- 10 km/s) at the 3
positions given above. We note that this  is larger than the 30 km/s
predicted from the relation of Terlevich \& Melnick (1981) between the line
absolute intensity and width, assuming log(F(H$\beta$))=40.62 (Mas-Hesse \&
Kunth 1997, in prep.) for Haro~2. 

\begin{figure}
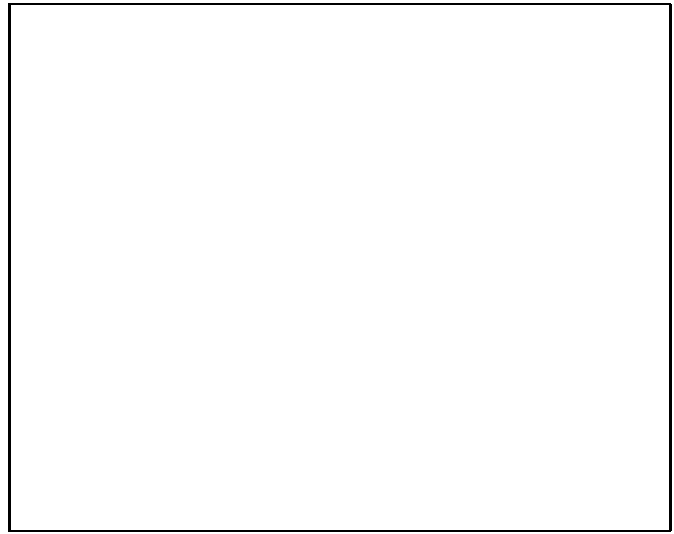

\picplace{7cm}
\caption[ ]{H$\alpha$ emission integrated over the central 7$\arcsec$ (CE) of Haro~2.
The unsmoothed observed H$\alpha$ emission profile is in solid line while
the gaussian fit is given in dashed lines. 
The excess intensity in the wings amounts to 2.2 \%  and 3.8 \% of the 
total H$\alpha$ line flux in the red and in the blue respectively.}
\label{fig:halpha}
\end{figure}

Moreover, the wings of the H$\alpha$ line (and also, the wings of the
nitrogen and sulfur lines) are broader than a simple gaussian profile (Fig.
\ref{fig:halpha}) and show a slight asymmetry. The flux excess in the
red and blue wings (as compared to a gaussian) is respectively 
around 2.2 \% and 3.8 \% of the total H$\alpha$ line flux.

\begin{figure}
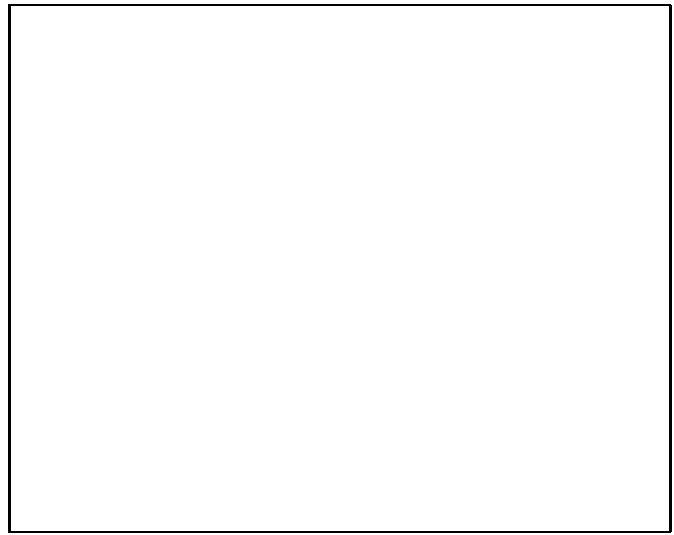

\picplace{7cm}
\caption[ ]{FWHM of the H$\alpha$ line versus position in Haro~2.
The represented error bars in the measurement of the FWHM amount to 0.3 \AA }
 \label{fig:fwhm}
\end{figure}

To assess the significance of this observed asymmetry in H$\alpha$, we have
compared it with the instrumental profile, by convolving a calibration line
of the same intensity with a gaussian of FWHM=2.7 \AA. The comparison of
both profiles shows that the wings of H$\alpha$ cannot be purely
instrumental but are intrinsic to the galaxy, confirming that the red wing
is about 35\% less intense than the blue wing (Fig. \ref{fig:halpha}).
We have also measured the position and the FWHM of the H$\alpha$ line along
the slit with an accuracy respectively better than 5 km/s for the position
and 0.3 \AA\ for the FWHM. We have noticed that the FWHM of H$\alpha$
increases from the central region to the external ones
(Fig. \ref{fig:fwhm}) and that simultaneously the relative contribution
of the broad wings to the total flux increases.

\begin{figure}
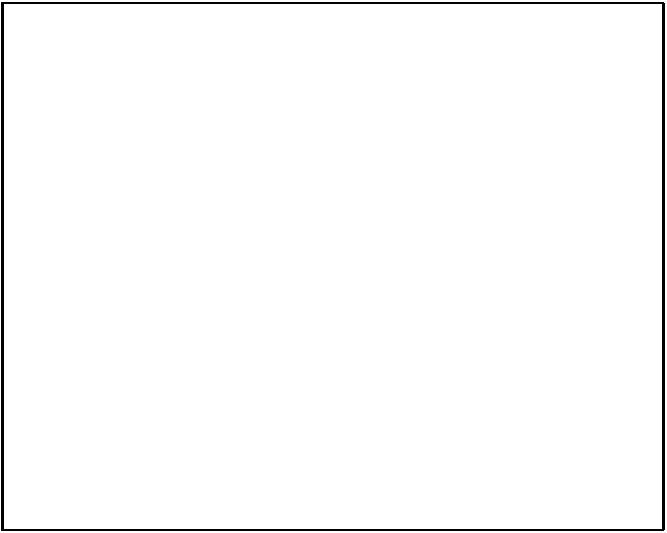

\picplace{7cm}
\caption[ ]{Rotation curve obtained from H$\alpha$ in Haro~2.
The error bars have been computed assuming an average 
uncertaintie of 5 km/s on the position of H$\alpha$ }
\label{fig:rotation}
\end{figure}

The rotation curve we have obtained shows two discontinuities at around
13$\arcsec$ from the center in both directions, as shown in
Fig.~\ref{fig:rotation}.  The velocity of the gas at these two
discontinuities is similar to the recession velocity of the nuclear region.

We have finally compared the profiles of the H$\alpha$ and the HST Ly$\alpha$
(Lequeux et al. 1995) emission lines, according to the following approach: 

\begin{enumerate}

\item  The spectra of the central part of Haro~2 have been extracted over
 6 pixels (2"), corresponding to the same aperture than the HST observations
 (Lequeux et al. 1995).

\item  We have blueshifted H$\alpha$ to the wavelength of the 
observed Ly$\alpha$, conserving its flux.

\item The H$\alpha$ emission line has been multiplied by 11.4, the 
theoretical Ly$\alpha$/H$\alpha$ ratio assuming case A in the recombination
theory and an electronic temperature of 10000 K (Osterbrock, 1989).

\item  The Ly$\alpha$/H$\alpha$ has been corrected from reddening 
effects, assuming a Small Magellanic Cloud extinction law with
E(B-V)=0.2. This has decreased the ratio to about 5. 

\item The transfer 
function for Ly$\alpha$ throught the interstellar medium has been
obtained by using Xvoigt (see Sect.~\ref{sect:stelpop}) .  

\item  Finally, the  blueshifted H$\alpha$ profile has been convolved 
with the transfer function for Ly$\alpha$ in order to obtain the intrinsic
Ly$\alpha$ profile one would expect if both lines were originated in the
same region under similar physical conditions.

\end{enumerate}  

\begin{figure}
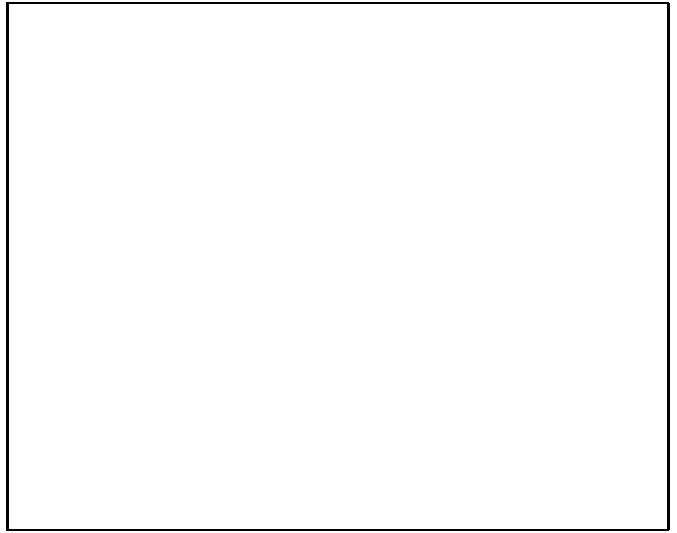

\picplace{7cm}
\caption[]{Comparison between the Ly$\alpha$ profile expected from
H$\alpha$ (ratios 5 in dashed lines and 11.4, in dotted lines) and that 
observed with HST (solid line).
The observed Ly$\alpha$ emission is always broader than the one expected from 
H$\alpha$} 
\label{fig:HalphaLYalpha3}
\end{figure} 

Comparing the observed Ly$\alpha$ profile with the predicted one 
(Fig.~\ref{fig:HalphaLYalpha3}) one can see that the observed Ly$\alpha$ line
appears to be, in all cases, much broader than expected. This implies the
existence of additional sources of Ly$\alpha$ photons at large
velocities.


\section{Analysis of the stellar population} \label{sect:stelpop}

Several authors (Larson 1986, Augarde \& Lequeux 1985, Rieke et al. 1993)
have suggested that, in active star formation regions, the production of
low mass stars should be hampered by the destruction of the molecular
clouds, in which low mass stars are formed, by the energy released by
massive stars and supernovae. This could result in a truncation of the
Initial Mass Function (IMF) at the low mass star end. Such a truncation can
have observable effects on the equivalent widths (EW) of the Balmer
absorption lines (Olofsson 1995).  As shown by various authors (e.g Cananzi
et al. 1993), the equivalent widths of the stellar Balmer lines is a
sensitive indicator of the IMF in very young populations.  We have thus
measured the absorption lines of the Balmer series in Haro~2, using two
different methods. First, we have used IRAF to fit two gaussians, one for
the emission line, one for the absorption line and second, we have used the
program Xvoigt\footnote{Xvoigt, Copyright 1994, David Mar} which allows to
adjust ``by eye'' a Voigtian absorption profile on the wings of the Balmer
lines. This program computes the observed absorption profile of different
atomic lines produced by a cloud for which we can fix the density, the
velocity dispersion and the redshift.  Both methods lead to the same
values, with differences smaller than 0.5 \AA. We adopted the values
measured with Xvoigt.  The results are summarized in Table
\ref{tab:balmer}. \\


    \begin{table}
    \small
    \caption [] {Measurement of the Balmer absorption lines.} 
    \label{tab:balmer}
    \begin{flushleft}
    \begin{tabular}{ccc}
    \hline
Line & Position & EW \\
    \hline
        & (1) & (2) \\
    \hline
H$\epsilon$	&	3970.1	&	4.6 $\pm$  0.5	\\	
H$\delta$	&	4101.7	&	4.7 $\pm$  0.5	\\
H$\gamma$	&	4340.4	&	4.1 $\pm$  1.0	\\
    \hline
    \end{tabular}
    \end{flushleft}
{\em Columns: }

(1) : Rest wavelength in \AA.\\
(2) : EW in \AA \ measured with Xvoigt. \\
    \end{table}

Fanelli et al. (1988) have analyzed the UV spectrum of Haro~2 using
optimized, non-evolutionary, stellar synthesis techniques. These authors
derived a strongly discontinuous luminosity function for this galaxy,
concluding that the present burst of star formation has been preceded by at
least two bursts, the most recent of which ended not more than 20 Myr ago.
Mas-Hesse \& Kunth (1997 in prep.) have also analyzed this galaxy applying
evolutionary synthesis techniques. 
They obtain an age of
4 -- 5~Myr for the present burst, with best fits obtained for a rather
steep ($\alpha \geq 3.0$) IMF (defined as $dN/dm=\phi(m)\propto m^{- \alpha}$)
and a nearly instantaneous star-formation episode. 
The models of Mas-Hesse \& Kunth (1997, in prep.), based on the fit of the UV
continuum, clearly underestimate the optical continuum. The stars formed in
the present burst, the only one to which the UV range is sensitive, account
indeed for not more than 60\% of the total optical emission,
indicating that an important fraction of the stellar population was formed
prior to the present burst. Kr\"uger et al. (1995) obtain indeed also a
good fit to the optical and near infrared continuum by assuming that a
5~Myr long starburst has taken place on a galaxy having formed stars
continuously, but at a
lower rate, during the last 15~Gyr. Finally, Loose \& Thuan (1986) detected a
change in stellar populations from O to F6 main sequence stars as radius
increases, concluding that Haro~2 is an elliptical galaxy with very active
star formation activity in its center. 

We have compared the observed Balmer absorption line equivalent widths with
the predictions made by Olofsson (1995), based on evolutionary models
computed assuming various truncations of the IMF.  To be consistent with
previous results, we have assumed a two-burst model with IMF slope $\alpha
= 2.5$.  The relative strength of each burst was chosen to reproduce the
total optical emission. No unique solution can be derived from the
calculation but we have found that in any case the observed equivalent
widths exclude models in which the IMF is truncated below 10
$M\odot$. Therefore, if the onset of a starburst inhibits the formation of
low mass stars, this result shows that the process should be effective only
at masses clearly below 10 $M\odot$, if any.

According to the results by Mas-Hesse \& Kunth (1997, in prep.), the mass 
transformed into stars in the mass interval between 2 and 120 $M\odot$ is 
rather high (more than 10$^6$ $M\odot$), and the burst is in a quite late 
phase (at around 4.5~Myr), so that the amount of kinetic energy released 
by such an episode (stellar winds, supernovae explosions) is large enough 
for having disrupted the interstellar medium, as we will discuss below.


\section{Discussion}

The excess in the velocity dispersion measured in the central region of
Haro~2 with respect to the correlation from Terlevich \& Melnick (1981) can
be explained by the combination of rotation and seeing effects. The
rotation curve shows a variation in the position of H$\alpha$ of 0.5 \AA \
(20 km/s) per 5-6''. Since the seeing was of this order, we would expect an
increase in the observed FWHM of the lines by about 20 km/s. It seems
therefore that the correlation given by these authors should flatten for
large galaxies in which the rotation introduces a measurable effect, while
it remains valid for smaller HII galaxies with no significant rotation.

\begin{figure}
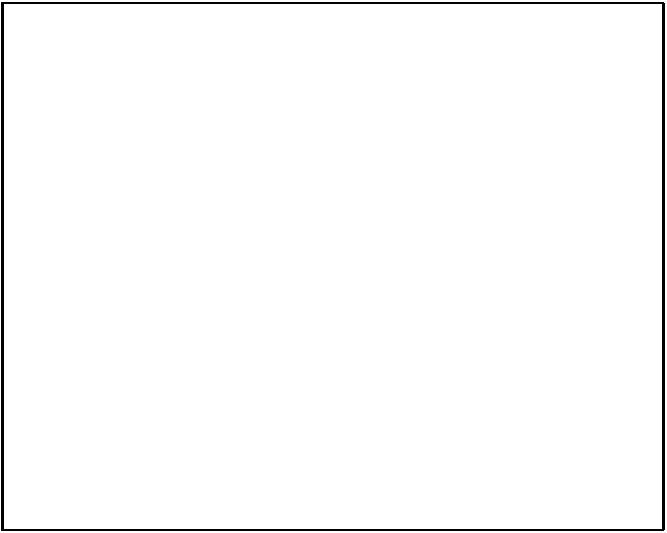

\picplace{7cm}
\caption[]{Line fitting to the observed H$\alpha$ line in the central (CE) 
region.
The three component gaussian fit is given by the dashed line while the
observed H$\alpha$ emission is represented by the solid line.
The three fitted gaussians are at 0, +200 and -200 km/s with respect to 
the center of the line.
The intensity contribution of the two components at +200 km/s and -200 km/s
is around 7\% of the central line.}
\label{fig:fit200kms}
\end{figure}

Lequeux et al. (1995) have shown evidences for the existence of an
expanding shell at 200 km/s in Haro~2.  We will discuss now how our optical
observations can be interpreted in this scenario, and will derive a simple
geometrical model for Haro~2.  As we have shown above, the H$\alpha$
emission profile shows two broadened wings.  By assuming that the total
H$\alpha$ line is due to the emission from a central HII region and to the
additional emission from a partially ionized shell expanding at 200 km/s,
we have attempted to fit the H$\alpha$ profile by the sum of three
gaussians with velocities of 0, +200, and -200 km/s with respect to the
center of the H$\alpha$ line (Fig.~\ref{fig:fit200kms}).  Our fit can
reproduce the broadening of the line profile, indicating that the wings are
consistent indeed with the emission coming from an expanding shell. The
total contribution of this shell to the total H$\alpha$ luminosity would be
close to 7\%. The residuals that are evident from the fit shown in
Fig.~\ref{fig:fit200kms} would be reduced by considering a more realistic 
convolution of
gaussians at velocities around $\pm$200 km/s, as expected in an expanding
shell, instead of single velocity values.

Finally, the broadening of the wings is asymmetric. This can be explained
if the broadening is indeed due to an expanding shell. Photons coming from
the receding part of the shell are likely to be absorbed by dust within the
HII region and the front part of the shell while those emitted by the
foreground part of the shell can escape directly.

The H$\alpha$ line intensity strongly decreases as we go away from the
center. The intensity scale between the center and a distance of 10 arcsec
or more is roughly a factor of 10. Over this distance the rotation curve
has been measured.  

At about 13'' from the center the velocity 
decreases to reach the value measured
in the central
region, before increasing again.
This suggests that we are
detecting emission from gas whose motion does not follow the galaxy
rotation and appears at the same velocity -the systemic velocity- than the
central region. We suggest that this gas is precisely the ionized part of
the expanding shell ejected by the central region. Since this effect starts
to appear at 3$\farcs$5 of the central region, we conclude that the radius
of the central HII region is around this value, which is consistent with
the estimation of 3$\arcsec$ by Lequeux et al. (1995). Therefore, at larger
distances the emission starts to be dominated by the shell.  We have
measured the contribution to the total H$\alpha$ luminosity of the two
regions at around 10$\arcsec$. This contribution is found to be around 6-7
\% , which is consistent with our previous estimation of the contribution
of the shell to H$\alpha$.  Moreover we argue that locations where the gas
emission appears at the same velocity than that of the nucleus correspond
to the ``border'' of the emission shell. This allows us to limit the radius
of the ionized shell to 13$\arcsec$ or 1.23 kpc (Ho=75 km/s/Mpc).  Marlowe
et al. (1995) have found bubbles with sizes of the same order. Using this
result and the expanding velocity of 200 km/s we can evaluate the
characteristic dynamical timescale of the shell to be 5-6 Myr.  This
timescale corresponds to the start of the present nuclear starburst,
estimated at 4-5 Myr by Mas-Hesse \& Kunth (1997, in prep.). 

We can estimate the kinetic energy of the neutral expanding gas that is
pushed out by the shell by 

$E_{k}( \rm HI \it )=\frac{1}{2}4 \pi r^{2} N( \rm HI \it) m_{p} v^{2}$, 

\noindent
where $r$ is the radius of the shell, \it N \rm (HI) the column density of the
neutral gas, $m_{p}$ the mass of the proton and $v$ the mean expansion
velocity.  Lequeux et al.  (1995) have evaluated the column density of the
neutral gas expanding with the shell to be $ N(\rm HI \it) = 7 \ 10^{19} \rm \
cm^{-2}$ on the line of sight. Adopting this value we find 
$ E_{k}( \rm HI \it )= 4.53 \ 10^{54} \rm \ ergs$ and a mass for the 
expanding gas of $6 \ 10^{6}$
$M\odot$. We have applied the relation of Marlowe et al. (1995) by using
the Castor et al. (1975) model which gives the evolution of the radius and
the expansion velocity of a bubble expanding adiabatically:

$r_{bubble}=1.0(\frac{dE}{dt})_{41}^{\frac{1}{5}} n_{0}^{-1/5} t_{7}^{3/5}
 \rm \ kpc$ \hfill (1)\

$v_{bubble}=62(\frac{dE}{dt})_{41}^{1/5} n_{0}^{-1/5} t_{7}^{-2/5}\ \rm\ 
km \ s^{-1}$ \hfill(2)

\noindent
where $(\frac{dE}{dt})_{41}$ is the kinetic energy injection rate in units
of $10^{41} \rm \ erg \ s^{-1}$, $ n_{0}$ is the number density in the
medium, and $t_{7}$ is the time since the expansion began in units of
$10^{7}$ yr.  The density of the medium is not known, and Marlowe et
al. (1995) adopt $n_{0}=0.3$. The kinetic energy injection rate for Haro~2
is $4.34 \ 10^{41} \rm \ erg \ s^{-1}$ at 6 Myr (Cervino et al. 1997, in prep.).
Using this value and adopting, rather arbitrarily, $n_{0}=0.3
\rm \ cm^{-3}$, we have evaluated the size and speed of the bubble. We find
$r_{bubble}=1.26 \ $ kpc and $ v_{bubble}= 130 \rm \ km \ s^{-1}$ for an age
of 6 Myr. Taking into account the inherent uncertainties, this is
consistent with our estimations for the size and the velocity. The
discrepancies with the measured values can be partially explained by the
expected asymmetry of the shell. No neutral gas at the systemic velocity
has been detected with HST in front of the central cluster, although we
know from radio observations that there are significant amounts of neutral
hydrogen surrounding Haro~2. We expect therefore that in the line of sight
we are detecting outflowing gas which is reaching the limits of the
cloud. On the other hand, the gas column density should be much larger in
the direction perpendicular to the line of sight, where we have measured
the size of the shell. We have therefore used the
precedent relations to constrain the density in both directions, using the
computed kinetic energy and  other observed parameters for a 6 Myr old
shell.  In formula (1), using the radius of 1.26 kpc, we have computed for
the direction perpendicular to the line of sight, a value of
$n_{0}=0.33 \ \rm cm^{-3}$ as an upper limit. 
In formula (2), the velocity 
of 200 km/s measured on the line of sight with the HST gives a value of
$n_{0}=0.03 \ \rm cm^{-3}$ as a lower limit in this direction.

We have mentioned that the FWHM of the H$\alpha$ line increases with
the distance to the center.
This means that as we are leaving the central region (along the slit), 
the contribution of the central HII region to H$\alpha$ decreases,
but not the one of the shell. This results in an increasing contribution of
the shell to H$\alpha$ and then to a broadening of this line due
to  the shell. The double peaked profile expected for the expanding shell  
alone is most probably smoothed by the poor
seeing. New observations around H$\alpha$ with high spatial and spectral
resolution under good seeing conditions may allow us to see this effect.

Comparison between H$\alpha$ and Ly$\alpha$ has revealed the existence of
additional sources of Ly$\alpha$ photons at high velocity. If the receding
ionized shell (or any other ionized region) were responsible for this
emission, it would also produce H$\alpha$ photons with the same kinematical
properties and showing therefore similar profiles. We interpret this extra
emission as photons backscattered by resonant diffusion on the receding
neutral part of shell, as suggested by Lequeux et al. (1995). Note that
these backscattered photons, diffused at +200 km/s, would not be absorbed
at all by the medium in the line of sight, which is at very different
velocities.

From the results we have discussed up to now, we can sketch a simple
geometrical scenario for Haro~2 inspired on several theoretical frameworks
(Castor et al. 1975: Weaver et al.  1977; Martin 1996; Tenorio-Tagle 1996):
A central starburst ionized region, at the systemic velocity, contains most
of the recent stars.  Since the HST spectra show the presence of OI and
SiIII absorptions at -200 km/s, we infer the existence of a partially
ionized expanding shell at this velocity, whose emission should contribute
to the observed wings of H$\alpha$. The inner part of the ionized region
could be constituted by the ejected materials from the central starburst
while the outer part would be formed by the ionized interstellar medium.  
The inner
gas is photoionized and perhaps partially ionized by shock-waves
propagating from the interfaces between the ejected and the interstellar
media (if the ionization produced by shocks is important, we should see
some typical emission lines, like [OI]6300\AA, which is not the
case. Nevertheless, these lines are expected to be weak and since the
contribution of the shell is very small they would remain below our level
of detection). Since Ly$\alpha$ emission from the central region at 0 km/s
is observed, there cannot be static HI gas on the line of sight. A
``champagne flow'' outcoming from the internal regions could be starting to
form here. On the other hand, we speculate that larger amounts of static
neutral gas could be present in the direction perpendicular to the line of
sight, being responsible for the bulk of HI radio emission detected at the
systemic velocity.  The neutral outer layers of the expanding shell would
be responsible for the absorption of Ly$\alpha$ photons detected with
HST. On the other hand, diffusion at the receding sections of the shell
would backscatter Ly$\alpha$ photons which would explain the excess of
emission detected in the red wing of this line.

This scenario exemplifies once again that the detectability of Ly$\alpha$
emission in starburst galaxies requires the combination of several effects,
as concluded by Kunth et al. (1996): 

\begin{itemize}

\item The presence of an ionized medium producing Ly$\alpha$ photons. 

\item The absence of neutral gas at the velocity of the Ly$\alpha$ emission
source in the line of sight, i.e., the presence of large gas flows in the
line of sight, not hidden by more external static, neutral clouds.  

\end{itemize}

\noindent

Therefore other effects suggested in the past, like dust abundance, 
metallicity, evolutionary state of the cluster, etc..., do not play the 
only role in the process. 
This explains why the very metal deficient and dust-free galaxy
IZw~18 shows nevertheless no Ly$\alpha$ in emission. IZw~18 is surrounded
indeed by large amounts of neutral HI gas at low velocity.  Ly$\alpha$
photons diffusion by neutral hydrogen would be produced independently on
the presence or absence of dust grains. If dust particles are abundant,
they would end absorbing the major part of the diffused photons. On the
other hand, if dust effects are not significant, a glow of leaking
Ly$\alpha$ photons should be detected around the whole neutral HI cloud
surrounding starburst galaxies. Maybe efficient detectors in the future
will be able to detect this Ly$\alpha$ glow.


\section{Conclusion}

We summarize the  main results of this work: 

\begin{itemize}

\item  We have confirmed the existence of a partially ionized 
shell expanding at 200 km/s in Haro2. The contribution of this shell
to the total H$\alpha$ luminosity is around 7\%. 

\item The size of the shell has been evaluated to
be 1.23 kpc and its age to 6 Myr, being apparently related 
to the last starburst event.

\item The shell is not affected by the rotation of the galaxy.  

\item  We confirm the Lequeux et al. (1995) hypothesis that 
backscattering is responsible for the broadening of the red wing of the
Ly$\alpha$ emission.

\item This starburst event probably has an IMF including stellar 
masses lower than 10 M$\odot$. \\

\end{itemize}

We have shown that long-slit high resolution spectroscopy around H$\alpha$
combined with Ly$\alpha$ spectroscopy is a powerful tool to study the
geometry and kinematic of the ISM. These observations of Haro~2 show that
Ly$\alpha$ photons can escape because of a particular structure and
kinematic of the ISM, as suggested by Kunth et al. (1996) and Giavalisco et
al. (1996).  Both factors strongly affect the profile of the Ly$\alpha$
emission line and become major drivers for its detectability. 


\begin{acknowledgements}
We want to thank S. Charlot, M. Cervino, M. Fioc, E. and R. Terlevich for
stimulating discussions. F.L. thanks the LAEFF-Madrid for his
hospitality during the period this work was performed. This work has 
benefited from the financial support of the Action Int\'egr\'ee
PICASSO. J.M.M.-H. has been partially supported by Spanish CYCIT grant
ESP95-0389-C02-02. 
\end{acknowledgements}

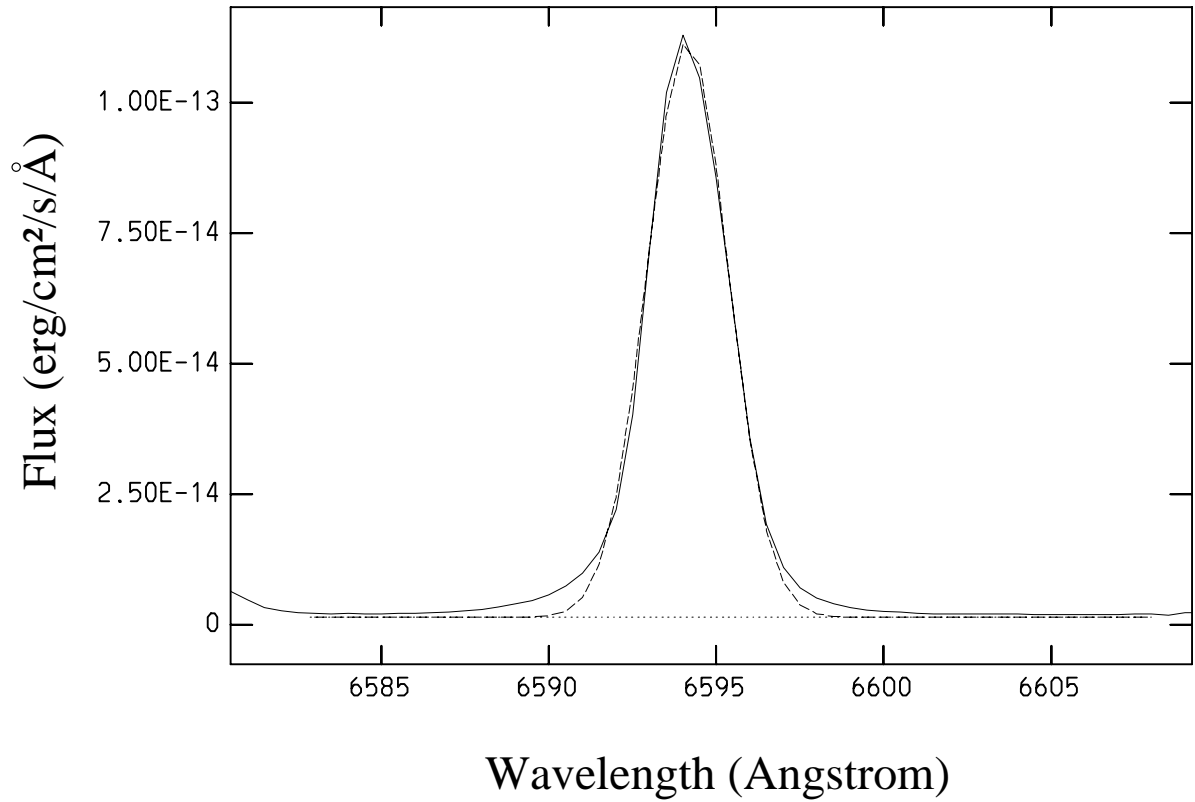
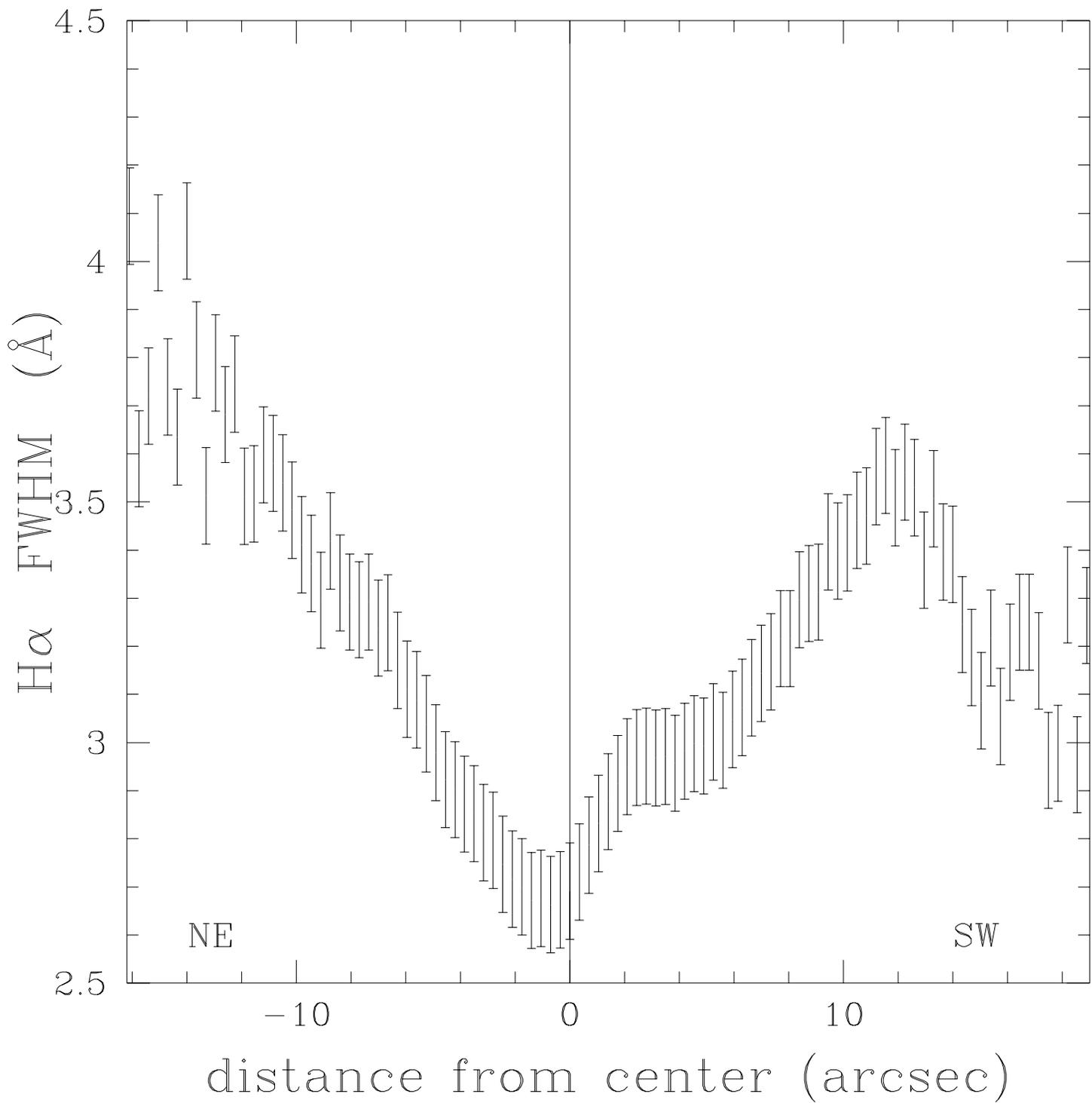
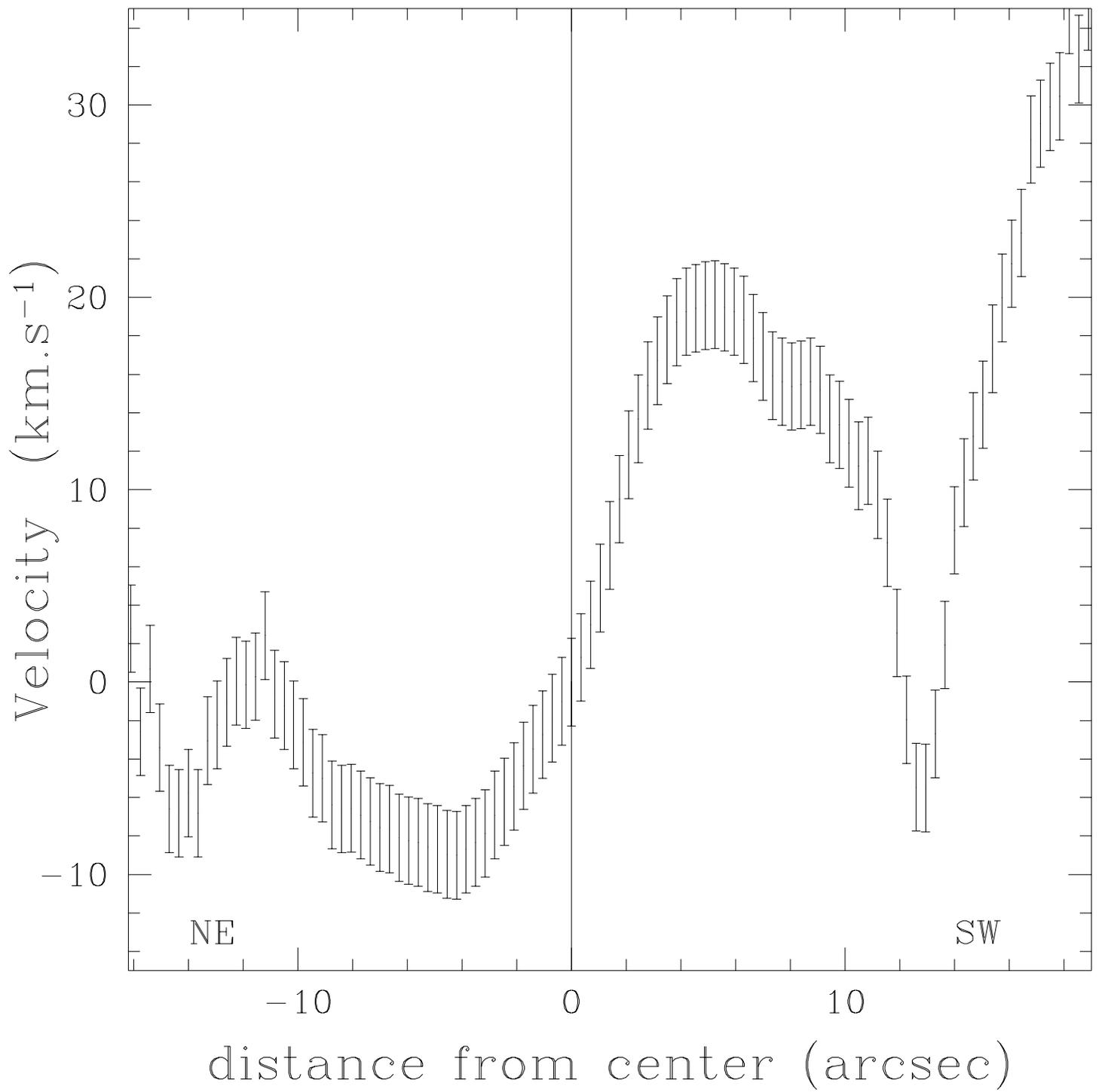
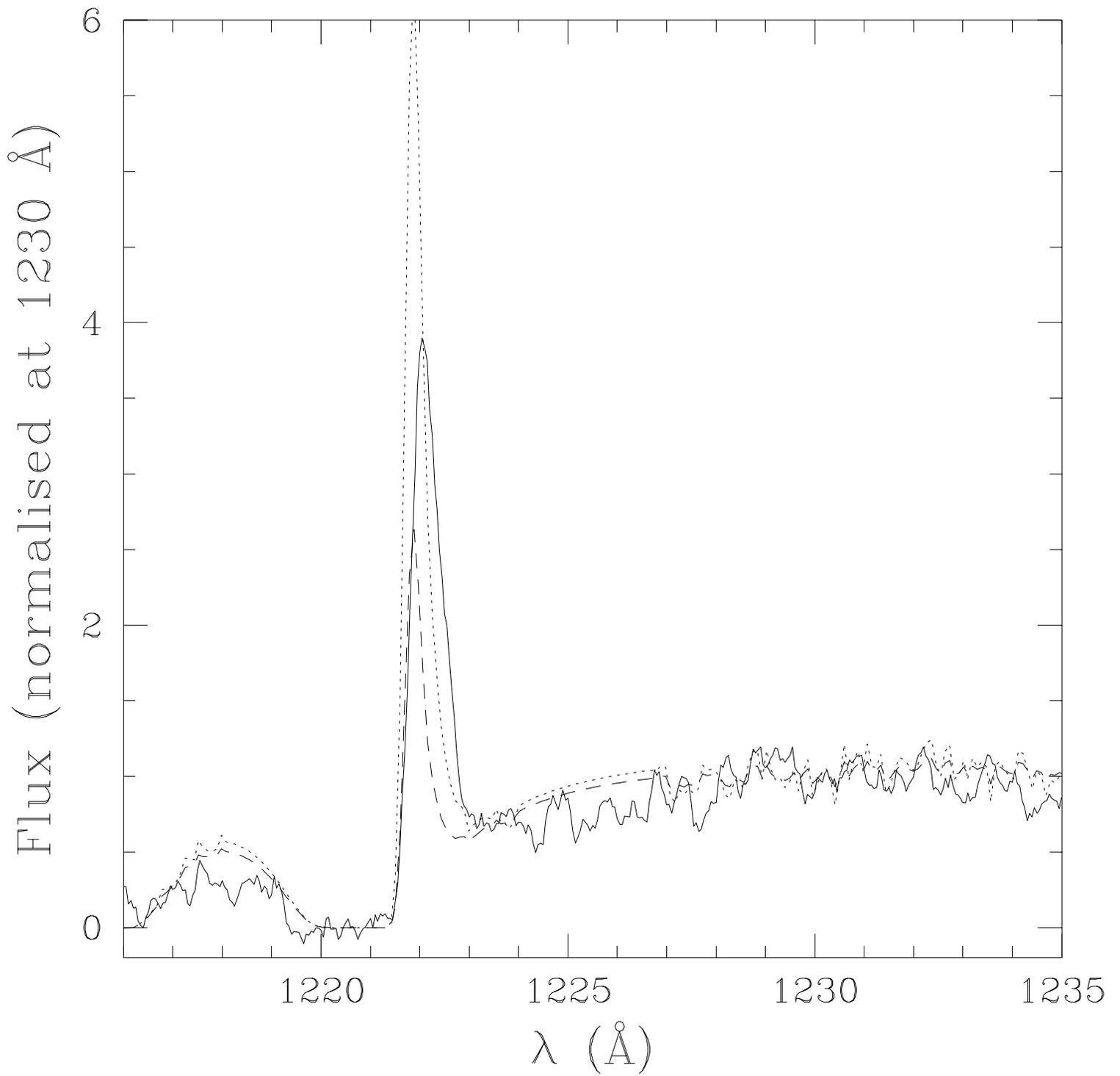
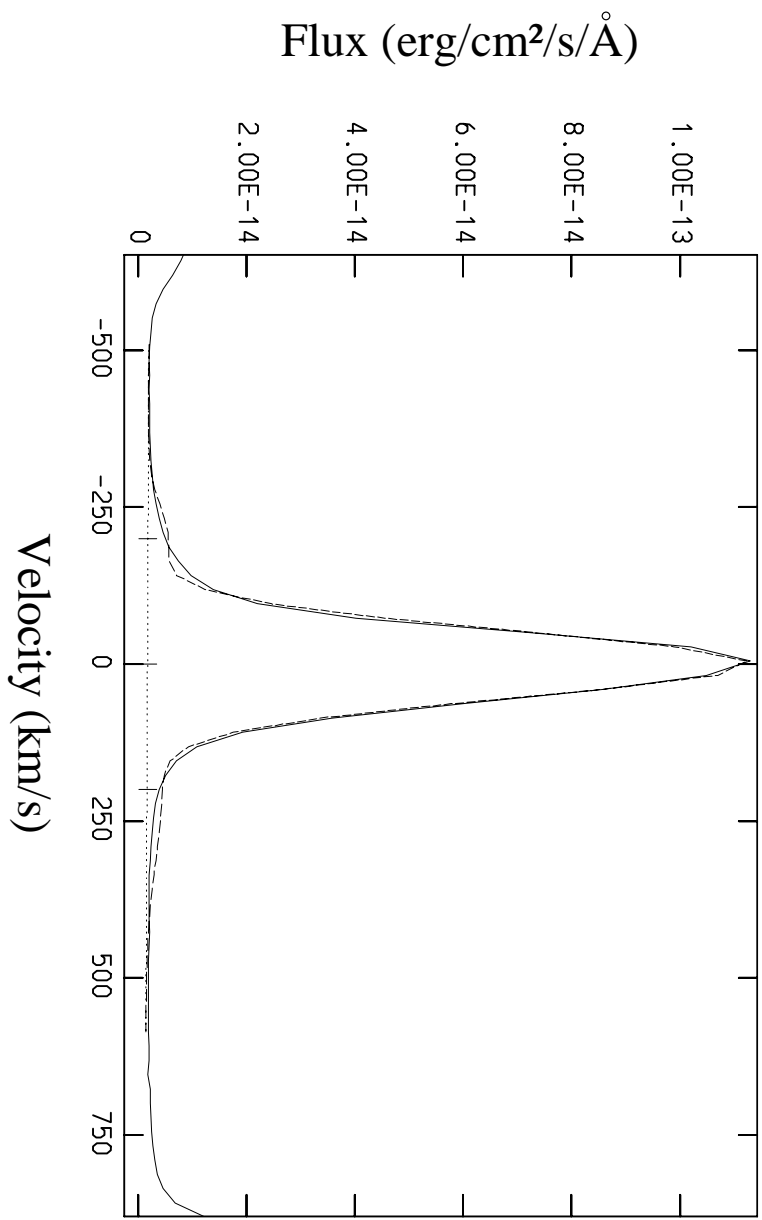
\end{document}